\documentclass{llncs}
\usepackage{amssymb}
\usepackage{graphicx,multirow,amsmath}
\usepackage[T1]{fontenc}

\usepackage{url}

\begin{document}

\title{A hierarchical cellular automaton model \\ of distributed traffic signal control}
\titlerunning{A hierarchical cellular automaton model}  
%
\author{Bart\l{}omiej P\l{}aczek}
\authorrunning{Bart\l{}omiej P\l{}aczek} 
%

%
\institute{Institute of Computer Science, University of Silesia, Katowice, Poland\\
\email{placzek.bartlomiej@gmail.com}}

\maketitle              

\begin{abstract}
This paper introduces a hierarchical cellular automaton (HCA) model for simulation of distributed self-organizing control of traffic signals at intersections in road network. The proposed HCA consists of three hierarchy levels that describe the movement of particular vehicles, occupancy of traffic lanes, and signal phases at intersections. Update rule of the HCA was designed to control traffic signals and minimize delays of vehicles in the road network. The introduced update rule takes into account states of cells from different hierarchy levels of the HCA that represent neighboring traffic lanes and intersections. Simulation experiments were conducted for a wide range of traffic conditions - from free flow to saturated traffic in two scenarios: Manhattan-like grid road network, and arterial road. Results of the simulations show that the proposed HCA-based traffic control strategy achieves better effectiveness in comparison with the state-of-the-art back pressure algorithm.
\end{abstract}
\section{Introduction}
Hierarchical cellular automata (HCA) are useful for modelling processes that operate at different scales within the same system \cite{bpcite:1}. This category of processes includes the distributed self-organizing control of traffic signals in road networks \cite{bpcite:2,bpcite:3}. For a comprehensive modelling of the distributed traffic signal control, it is necessary to consider the fine-scale dynamics of vehicles, adaptation of signals to local traffic situation at the intermediate scale, and coordination between signalized intersections at the coarse scale. In this paper a HCA-based model is introduced that handles the above-mentioned requirements. 

HCA is composed of cell levels that establish a hierarchy of cellular spaces. A higher level in HCA corresponds to cellular space with fewer cells, and represents a physical system with fewer components \cite{bpcite:1}. A neighborhood between cells from different levels in HCA is referred to as the interlevel neighborhood. In contrast, intralevel neighborhood connects cells within a single level of abstraction. 

Update rule in HCA is used to modify the states of cells at successive time steps of simulation. The update rule can be defined over the set of states produced by both the intralevel neighborhood and the interlevel neighborhood. The intralevel neighborhoods are used for information exchange at specific level in the hierarchy. For example, in the model of distributed traffic signal control, the coordination between intersections is modeled using information flow via intralevel neighborhoods that comprise a set of cells representing the neighboring intersections. Interlevel neighborhoods are used to transfer state information up and down the hierarchy, for instance between the cells representing traffic lanes and intersection.

The objective of the considered traffic signal control is to improve utilization of the existing road infrastructure, increase its capacity, and decrease delays of vehicles. According to the distributed control methods \cite{bpcite:2,bpcite:3}, traffic signals at each intersection in a road network are controlled independently by using an algorithm, which takes control decisions on the basis of real-time traffic data obtained from local measurements. These input data can describe current traffic situation at the considered intersection as well as status of traffic signals at neighbouring intersections. Output of the algorithm is a control decision that determines which traffic streams at the intersection should get a green signal for a subsequent time step.

The distributed control of traffic signals enables effective coordination of the vehicular traffic flows at the network level, without using any central controller. This approach has received a considerable attention from both industry and academia due to its important advantages, such as: cost-effectiveness, robustness, scalability, adaptivity, and flexibility. The distributed control can respond to instantaneous traffic demands without using any predefined signalization schedules that are based on average traffic parameters, e.g., average velocity. The results available in the literature show that this strategy enables improved effectiveness when compared with traditional traffic signal control algorithms \cite{bpcite:4,bpcite:5}. The distributed control strategy is particularly advantageous from the perspective of new traffic signal control systems based on vehicular sensor networks (VSNs) \cite{bpcite:6}. Such systems collect detailed input data from sensors installed in vehicles by using the wireless communication.

The paper is organized as follows. Section 2 gives a short overview of existing HCA applications and describes previous works related to the distributed traffic signal control. The proposed HCA model of traffic signal system is presented in Section 3. Section 4 includes experimental results and a comparison of the introduced method with state-of-the-art approach. Conclusions and future research directions are given in Section 5.

\section{Related Works and Contribution}

So far several different applications of HCA have been reported in the related literature. Dunn has proposed a HCA framework for modelling isotropic propagation and coupled propagation processes in heterogeneous spatial systems \cite{bpcite:1}. It was suggested that the above-mentioned framework can be useful for process modelling in landscape ecology and biosecurity. This approach was used to develop models of wildfire spread \cite{bpcite:7} and invasive plant spread \cite{bpcite:8}. In \cite{bpcite:9} a generalized definition of spatial hierarchy was used to introduce a HCA, which enables developing spatially hierarchical models on both the plane and the sphere. This work provides a basis for modelling various processes on the surface of the Earth.

Other application areas of HCA include electronics and computer vision. Sikdar et al. have employed a HCA to design a test pattern generator for testing very large scale integration circuits specified in hierarchical structural description \cite{bpcite:10}. The HCA in that study was designed based on the theory of the extension field. In \cite{bpcite:11} the Authors have introduced a HCA, which consists of single-layer cellular automata and cuboid cellular automata. That HCA was used to detect salient objects, i.e., the most important parts of an image.

This paper introduces a HCA for modelling distributed self-organizing control of traffic signals at intersections in road network. The proposed HCA consists of three hierarchy levels that describe the movement of particular vehicles, occupancy of traffic lanes, and signal phases at intersections. Update rule of the HCA reflects the operation of distributed traffic control algorithm. The proposed model can be used to evaluate effectiveness of available algorithms and to improve them by introducing control rules, which take into account states of cells from different hierarchy levels that deliver the traffic information with various resolutions.

The classical cellular automata have been implemented in several research works related to distributed self-organizing traffic signal control. A traffic model based on elementary cellular automata was used in \cite{bpcite:12} to study the effectiveness of the distributed self-organizing control. Another work \cite{bpcite:13} has explored the self-organizing ability of traffic signals by using a macroscopic two-dimensional cellular automata model of an urban traffic signal system. In that model each intersection is regarded as a cell and the "pressure" of vehicular flow is considered as a state of the cell.

Szklarski \cite{bpcite:4} has used the classical cellular automata to simulate a distributed traffic control system, which takes into account traffic stream priorities that correspond to expected increase of vehicle delay. In \cite{bpcite:6} this method was extended by introducing a traffic model, which combines interval arithmetic with cellular automata to predict delays of individual vehicles in a short time horizon and evaluate uncertainty of this prediction.

A cellular automata model of distributed traffic signal system was implemented in \cite{bpcite:3} to enable evolutionary optimization of control rules. It was shown that the cellular automata facilitate a fast simulation-based evaluation of the fitness function, which guides the evolution of control rules towards an optimal solution.
 
In this paper the new HCA model is proposed and utilized for improving the effectiveness of distributed traffic control based on back-pressure method. Initially, the back-pressure algorithm was intended for routing in wireless computer networks to provide maximum throughput under the assumptions that all links in the network have infinite capacities. This concept was then adapted to urban road networks for distributed signal control \cite{bpcite:2}. According to the back-pressure algorithm, a higher priority is assigned to traffic streams with high upstream pressure and low downstream pressure. The pressures correspond to numbers of vehicles and the priority is proportional to so-called differential traffic backlog, i.e., difference of vehicle numbers in inbound and outbound traffic lanes at the intersection.

\section{Proposed Model}

This section introduces a hierarchical cellular automata (HCA) model, which enables simulation of distributed control of traffic signals at intersections in road network. The proposed cellular automaton consists of three levels that describe the movement of particular vehicles, occupancy of traffic lanes, and signal phases at intersections (Fig. 1). Update rules of the introduced HCA are designed to control traffic signals and minimize delays of vehicles in the road network.

\begin{figure}
\centering
\includegraphics[height=5cm]{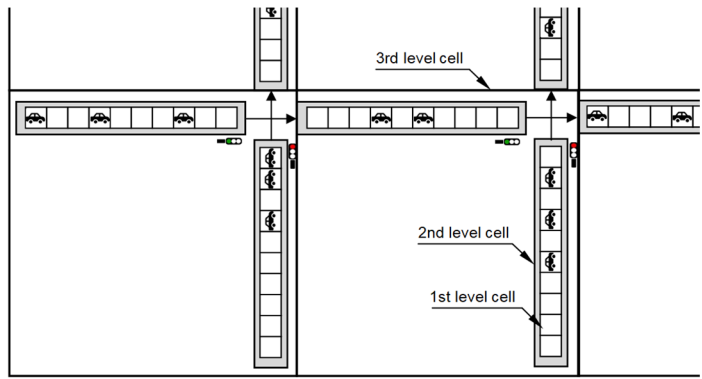}
\caption{Structure of the proposed hierarchical cellular automaton.}
\label{bp:fig1}
\end{figure}

\subsection{Vehicle level}
The lowest (first) hierarchy level of the proposed HCA includes information about particular vehicles, i.e., their speeds and positions. In order to simulate the flow of vehicles, traffic lanes are divided into cells of equal length. Positions of vehicles in traffic lane are indicated by occupied cells. Speed is determined as a number of cells that vehicle passes during one time step. These assumptions are consistent with those of the state-of-the-art cellular automata models of road traffic \cite{bpcite:14}. 

State of cell $k$ at the first hierarchy level is represented by variable $\dot{\sigma}_k$. For empty cell the state $\dot{\sigma}_k$ equals -1. If $\dot{\sigma}_k$ is above -1 then cell $k$ is occupied by a vehicle, and $\dot{\sigma}_k > -1$ denotes speed of the vehicle (in cell per time step). The maximum speed is determined by parameter $v_{\max}$.

In this study, the state updating operation for first level cells is based on a rule, which was introduced by Brockfeld et al. \cite{bpcite:15} for a stochastic cellular automata model of Manhattan-like road network with signalized intersections. According to this update rule, at each time step the new speeds and positions of vehicles are calculated as follows:
\begin{enumerate}
\item vehicle acceleration: $v_n \leftarrow \min(v_{n + 1}, v_{\max})$,
\item breaking due to other vehicles or traffic signals: 
\begin{itemize}
\item if the nearest traffic signal ahead is red then $v_n \leftarrow \min(v_n, d_{n - 1}, s_n - 1)$,
\item  if the nearest traffic signal ahead is green then $v_n \leftarrow \min(v_n, d_n - 1)$,
\end{itemize}
\item speed randomization: $v_n \leftarrow \max(v_n - 1, 0)$ with probability $p$,
\item vehicle movement: $k_n \leftarrow k_n + v_n$, if $k_n > k_n^{dest}$ then remove vehicle $n$,
\end{enumerate}
where: $k_n$ is location (cell) of vehicle $n$, $v_n$ is speed of vehicle $n$, $d_n$ is distance between vehicle $n$ and the next vehicle ahead, $s_n$ denotes distance between vehicle $n$ and the nearest signal ahead, $k_n^{dest}$ denotes destination cell for vehicle $n$. The distances are expressed in cells. It should be also noted that $\dot{\sigma}_k=v_n$ if  $k=k_n$.
The above rule is applied for all the vehicles at the same time. Operation 1 represents driver tendency to drive as fast as possible, operation 2 is necessary to avoid collisions, and operation 3 introduces random driving behavior. Finally, in operation 4 the vehicles are moved according to the new velocity calculated in steps 1-3.

\subsection{Traffic lane level}

The second hierarchy level of the proposed HCA includes cells that represent traffic lanes at signalised intersections. It is assumed that a traffic signal is localized at the end of each traffic lane represented by the second-level cell. State of the cell for lane $l$ is defined as triple:
\begin{equation}
\ddot{\sigma}_l=(o_l,\delta_l,\gamma_l),
\end{equation}
where $o_l$ is occupancy of lane $l$, $\delta_l$ denotes differential traffic backlog, which is used to control traffic signals in accordance with the backpressure method \cite{bpcite:2}, and $\gamma_l$ determines indication of traffic signal at the end of lane $l$ ($\gamma_l=1$ if signal is green,  $\gamma_l=0$ if signal is red).

Interlevel neighbourhood of cell $l$ $(\ddot{N}_l^*)$ is a sequence of the first-level cells that correspond to segments of traffic lane $l$: $\ddot{N}_l^*=(k,k+1,\ldots,k+m)$. Occupancy $o_l$ is determined as the number of non-empty cells in $\ddot{N}_l^*$. 

Intralevel neighbourhood $\ddot{N_l}$ is a set of the second-level cells $l_e$, such that a vehicle, which enters an intersection through lane represented by cell $l$, can exit this intersection through lane represented by cell $l_e \in \ddot{N_l}$. 

At each time step of the cellular automaton evolution, cells at second hierarchy level are updated using the following rule:
\begin{enumerate}
\item occupancy calculation: $o_l=|\{k:k \in \ddot{N}_l^* \wedge \dot{\sigma}_k > -1\}|$ ,
\item differential backlog quantity calculation: $\delta_l=\sum_{l_e \in \ddot{N_l}} [\omega_{l,l_e} \cdot (o_l-o_{l_e})]$,
\end{enumerate}
where $|\cdot|$  denotes cardinality of set, and $\omega_{l,l_e }$ is probability that vehicle entering intersection through cell $l$ will exit through cell $l_e$ ($\omega_{l,l_e}$ is evaluated based on historical traffic data). The above rule is applied in parallel for all the second-level cells, i.e., traffic lanes.

Additional update of the cells at second hierarchy level is performed after actualization of third-level cells. This additional update consists in setting the signal indication $\gamma_l$ to match the state of intersection signal phase, which is determined by the upper-level (parent) cell.

\subsection{Intersection level}

The cells at the highest (third) level of hierarchy represent signalised intersections in the road network. State of cell $i$ at the intersection level is defined as pair:
\begin{equation}
\dddot{\sigma}_i=(\pi_i,\tau_i),
\end{equation}
where $\pi_i$ denotes current active phase of traffic signals at intersection $i$, and $\tau_i$ is time elapsed from activation of phase $\pi_i$. Active signal phase specifies one or more inbound traffic lanes at the intersection that currently receive the right of way (green signal). For the grid road network presented in Fig. 1 two signal phases are available: in the first phase green signal is provided for vehicles approaching from the left, and in the second phase green signal is given for vehicles approaching from the bottom. It was assumed that each possible signal phase $\pi$ for a given intersection can be determined as a set of traffic lanes that get green light, i.e., $\pi=\{l\}$.

Intralevel neighborhood of the third-level cell $i$ $(\dddot{N}_i)$ includes cells that represent the nearest, directly connected intersections from which vehicles approach intersection $i$. Second-level cells ($l$) that represent lanes for inbound traffic at intersection $i$ belong to the interlevel neighbourhood of cell $i$ ($\dddot{N}_i^*$). 

State of each third-level cell is updated in successive time steps using the following rule: 
\begin{enumerate}
\item signal phase selection: $\pi_i=\arg\max_\pi\{\rho^*(\pi)+\alpha\cdot\rho(\pi)\}$,
\item if $\pi_i$ was changed then $\tau_i=0$ else $\tau_i\leftarrow\tau_i+1$.
\end{enumerate}

The signal phase selection (operation 1) is performed by taking into account combined priorities of the available phases ($\pi$). Two types of phase priority are considered in the proposed approach. Priority of the first type $\rho^*(\pi)$ is calculated based on the differential traffic backlog $(\delta_l)$ of second-level cells $l$ that belong to the interlevel neighbourhood $\dddot{N}_i^*$ (note that  $\pi \subset \dddot{N}_i^*$ ):
\begin{equation}
\rho^*(\pi)=\sum_{l\in\pi}\delta_l,
\end{equation}
Thus, when using the first-type priority, the signal phase with the highest total differential traffic backlog is preferred. This assumption is consistent with the concept of backpressure traffic control \cite{bpcite:2}. It should be noted that priority $\rho^*(\pi)$ takes into account state of inbound and outbound traffic lanes, while ignores the states of neighboring intersections.

State of traffic signals at neighboring intersections $i_\eta\in\dddot{N}_i$ is taken into account by the phase priority of second type $(\rho(\pi))$. Definition of this priority aims at coordinating the signal phases of neighboring intersections in order to strengthen the effect of so-called green wave. It means that vehicles, which exit a neighboring intersection $i_\eta$ and approach intersection $i$, should get green signal on time to avoid stopping at intersection $i$. Let time $(i_\eta ,i)$ denote the minimum time, which is necessary for the vehicles to move from intersections $i_\eta$ to intersection $i$. The first vehicle can enter intersection $i$ when $\tau_{i_\eta} \geq time(i_\eta ,i)$. Let us further assume that signal phases $\pi_{i_\eta}$ and $\pi$ have to be activated at intersections $i_η$ and $i$, respectively, in order to let the vehicles pass through these intersections. Then, the following function takes positive value when the vehicles approaching from intersection $i_η$ can enter intersection $i$ and thus signal phase $\pi$ should be activated:
\begin{equation}
f(i_\eta,\pi)=\left\{\begin{array}{ccc}
\tau_{i_\eta}-time(i_\eta,i) & : & \pi_{i_\eta} \sim \pi \\ 
-\infty  & : & \pi_{i_\eta} \not\sim \pi\\
\end{array}\right.
\end{equation}
Expression $\pi_{i_\eta}\sim\pi$ in (4) denotes that the vehicles leaving intersection $i_\eta$ during signal phase $\pi_{i_\eta}$ will get green signal at intersection $i$ if phase $\pi$ is active. 
The second-type priority of signal phase $\pi$ is evaluated based on function (4):
\begin{equation}
\rho(\pi)=\max_{i_\eta \in \dddot{N}_i}\{f(i_\eta,\pi)\}
\end{equation}
    
The proposed update rule for the intersection-level cells reflects operation of a traffic control system, which manages traffic signals. This update rule takes into account the information from two hierarchy levels of cellular automaton. Parameter $\alpha$ of this rule was introduced to adjust the impact of neighboring cells at the 3-rd hierarchy level. If $\alpha=0$ then the updated state of 3-rd level cells is determined only by the second level cells from the interlevel neighborhood, and the control of traffic signals is executed in accordance with the state-of-the-art backpressure algorithm. In case of $\alpha>0$ the changes of traffic signals are also influenced by state of the intralevel neighborhood, i.e., state of traffic signals at adjacent intersections.
After update of third-level cells, a relevant modification of signal indications ($\gamma_l$) for second-level cells is introduced. At each time step (1 second), new states of all cells in the proposed HCA are determined during several update operations. These operations can be summarized as follows: 
\begin{enumerate}
\item update of first-level cells (vehicle positions and speeds),
\item update of second-level cells (lane occupancies, differential traffic backlogs),
\item update of third-level cells (signal phases, phase durations),
\item update of second-level cells (signal indications).
\end{enumerate}

\section{Experiments}

The proposed HCA model was implemented for simulation of distributed traffic signal control. Simulation experiments were conducted for two scenarios: Manhattan-like grid road network, and arterial road. During simulations, total stop delay experienced by vehicles at intersections was counted to evaluate effectiveness of traffic signal control. Effectiveness of the introduced traffic control strategy, which utilizes information available on different hierarchy levels of the HCA, was compared with that of state-of-the-art back pressure algorithm.

For the first simulation scenario, topology of the simulated road network is a square lattice of 8 unidirectional roads with 16 signalized intersections (Fig. 2 a). In second scenario an arterial road with 4 signalized intersections is considered (Fig. 2 b). For both scenarios the distance between intersections is of 300 m, which corresponds to 40 cells. The maximum velocity parameter $v_{\max}$ was set to 2 cells per second (54 km/h) and the braking probability $p$ was 0.2.

Intensity of the simulated traffic flow is determined by parameter $q$ in vehicles per second (vehs/s). In case of scenario 1 this parameter refers to eight traffic streams entering the road network. For scenario 2, parameter $q$ determines the intensity of traffic in arterial road, while the intensity for side roads was set to 0.02 vehs/s. At each time step of simulation, vehicles were randomly generated with a probability equal to the intensity parameter. As a result, binomially distributed traffic flows have been obtained that approximate Poisson distribution. Each simulation run was conducted for 36000 time steps with constant traffic intensity ($q$).

\begin{figure}
\centering
\includegraphics[width=9cm]{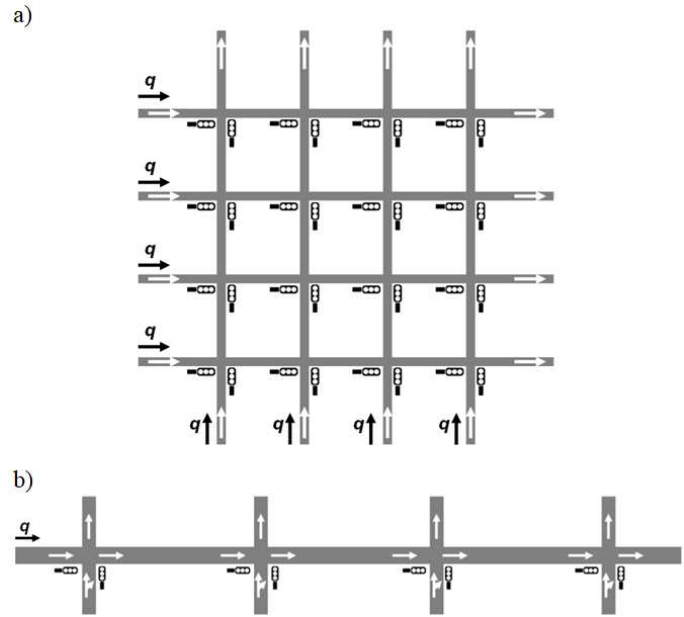}
\caption{Simulation scenarios: a) grid road network (Scenario 1), b) arterial road (Scenario 2).}
\label{bp:fig2}
\end{figure}
 
A first series of simulations was performed to examine the impact of parameter $\alpha$ on total stop delay of vehicles for the proposed HCA. Parameter $\alpha$ was changed in steps of 0.1. For each value of $\alpha$ the simulation was executed 50 times. Single run of the simulation corresponds to 3600 time steps, i.e., 1 hour. For this period the total stop delay of vehicles was counted. The results presented in Fig. 3 are averaged for 50 simulation runs. These results show that for both simulation scenarios and for all considered traffic intensities the minimum delay can be obtained if $\alpha$ is above 0. It should be noted that for $\alpha=0$ the standard back pressure traffic control algorithm is used. In case of Scenario 1 (grid network), the lowest delay was observed for $\alpha$ between 0.8 and 1.5. For Scenario 2 (arterial road) the best results were achieved with parameter $\alpha$ between 0.2 and 0.3. It can be also observed that the optimal $\alpha$ value (for which delay is minimized) increases for higher traffic intensities, especially in Scenario 1. 

The above observations confirm that the effectiveness of traffic signal control can be improved by making the control decisions based on sates of both the lane-level cells and the intersection-level cells. The effectiveness is decreased if states of the neighboring cells at the intersection level are ignored ($\alpha=0$). On the other hand, for high $\alpha$ values the influence of lane-level cells on control decision is marginalized and the decisions are mainly made on the basis of states of neighboring intersection-level cells. In this case, the delay of vehicles is also significantly increased. Thus, by selecting the value of $\alpha$, one can adjust the influence of the intersection-level and the lane-level cells on control decisions in order to find the optimal solution, for which the delay is minimized.

\begin{figure}
\centering
\includegraphics[width=12cm]{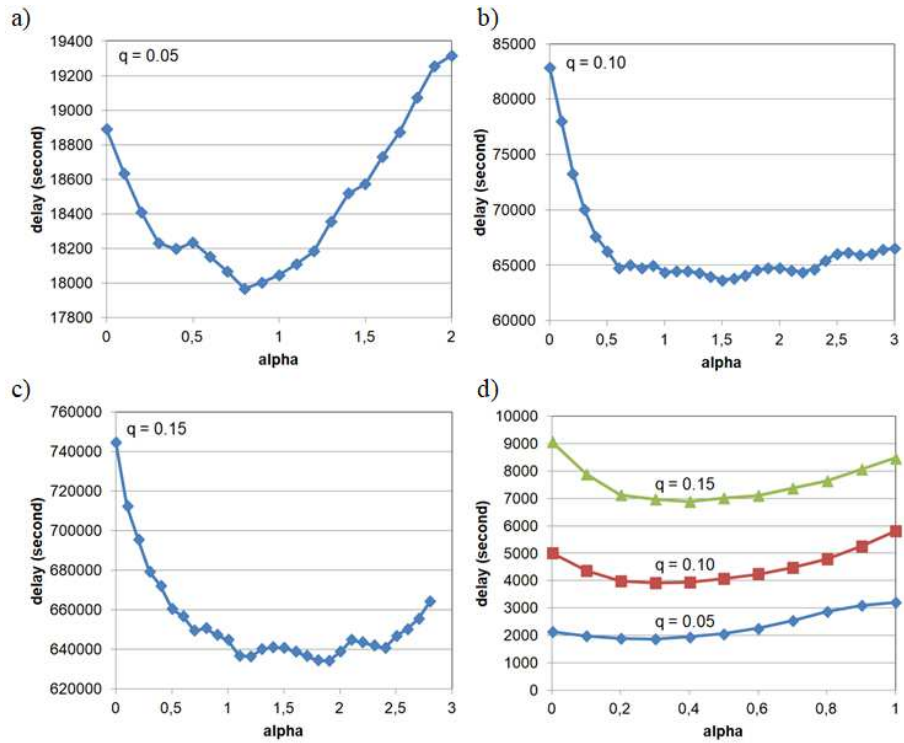}
\caption{Fig. 3. Impact of parameter alpha on total stop delay of vehicles for the proposed hierarchical cellular automaton: a)-c) Scenario 1, d) Scenario 2.}
\label{bp:fig3}
\end{figure}
   
Further analysis was performed to compare the effectiveness of traffic control for the proposed HCA approach and state-of-the-art back pressure strategy. Results of this analysis are presented in Fig. 4 for Scenario 1 and in Fig. 5 for Scenario 2.  Figs. 4 and 5 show the average delay values obtained for different traffic intensities. For each considered intensity value, the simulation was run 50 times. The columns in charts depict mean delay values for the 40 simulation runs. The error bars represent standard deviations. The considered intensity values (between 0.05 and 0.15 vehs/s) correspond to a wide range of traffic conditions - from free flow to saturated traffic. In both scenarios, for each setting of the intensity parameter the proposed HCA-based traffic control strategy achieves lower delay that the back pressure algorithm. On average the total delay of vehicles was reduced by 16\% in Scenario 1 and 20\% in Scenario 2.

\begin{figure}
\centering
\includegraphics[height=5cm]{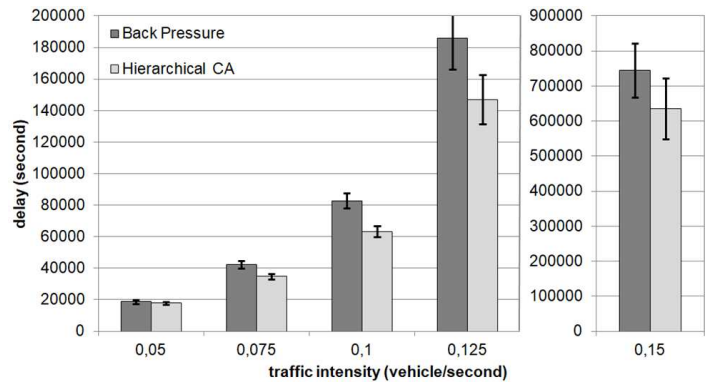}
\caption{Total stop delay of vehicles for the compared traffic control strategies (Scenario 1).}
\label{bp:fig4}
\end{figure}

\begin{figure}
\centering
\includegraphics[height=5cm]{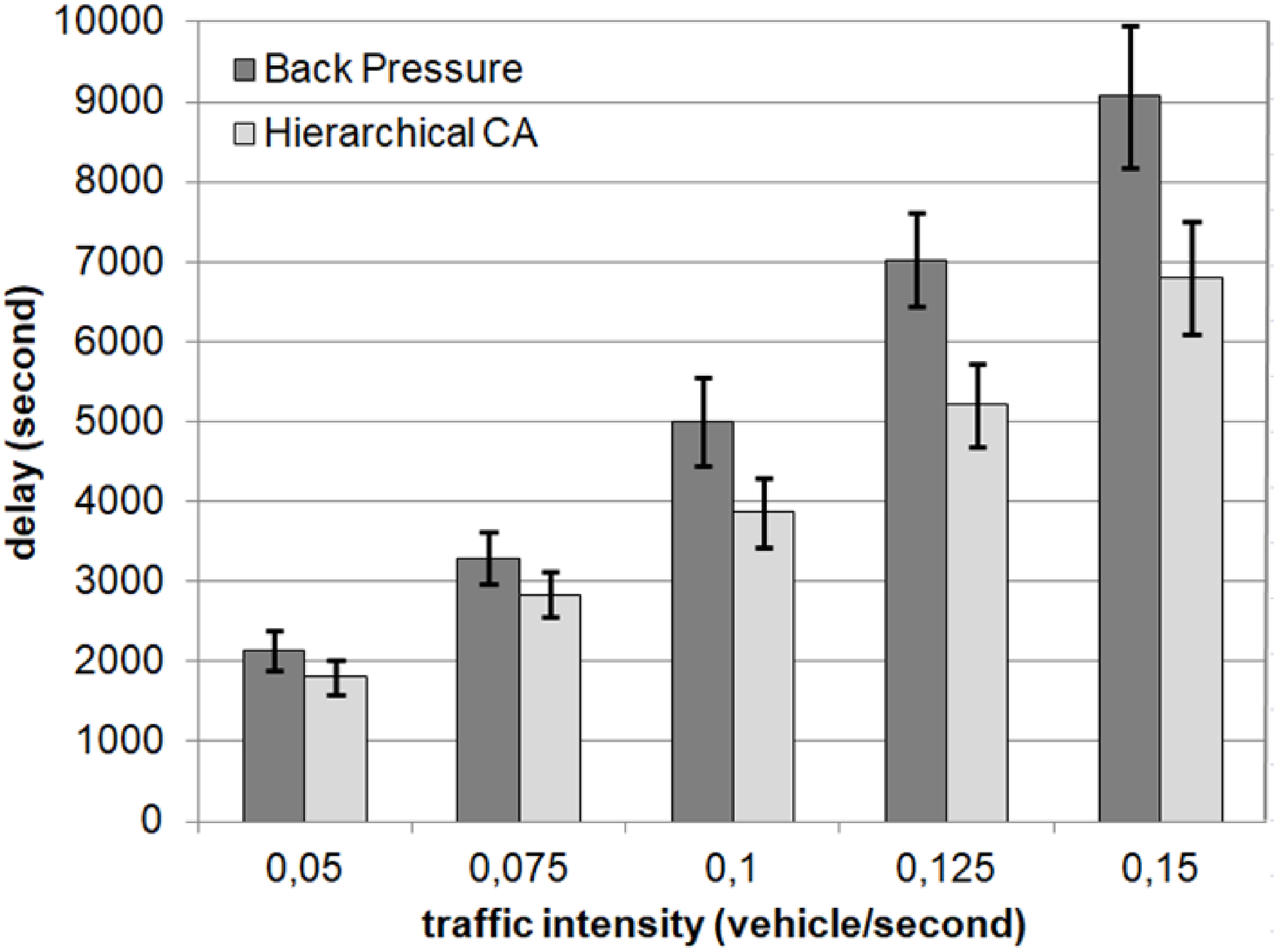}
\caption{Total stop delay of vehicles for the compared traffic control strategies (Scenario 2).}
\label{bp:fig5}
\end{figure}

\section{Conclusion}

In this paper the concept of HCA was adapted to simulate the operation of distributed traffic control system. Potential applications of the proposed HCA-based model include evaluation and optimization of existing signal control algorithms and development of new ones. The traffic signal control algorithm can be represented in HCA by update rules. An interesting feature of the proposed approach is the possibility of defining update rules that manage traffic signals based on the multi-resolution description of current traffic situation, which is provided by the HCA. An example of such update rule was presented in this study. The introduced update rule takes into account states of cells from different hierarchy levels that correspond to occupancies of traffic lanes in a given intersection and phases of traffic signals at neighbouring intersections. Simulation experiments show that this update rule achieves lower vehicle delay than the state-of-the-art back pressure algorithm. An interesting topic for future studies is the development of update rules that would take into account different criteria related to vehicle delay, traffic safety, and emissions. Another important task for further research is to verify the proposed approach in more realistic scenarios, with different examples of real-world road networks.

\section*{Acknowledgement}
This publication was supported by the Polish National Science Centre (Decision no. DEC-2017/01/X/ST6/00335).

%
%

\end{document}